\documentclass[a4wide,twocolumn]{article}

\usepackage{graphicx} 
\usepackage{url}      
\usepackage{listings} 

\usepackage{authblk}
\usepackage[square,numbers]{natbib}

\title{SuperCode: Sustainability PER AI-driven CO-DEsign}
\author[1,2]{P. Chris Broekema\thanks{Both authors contributed equally to this paper.}}

\author[2]{Rob V. van Nieuwpoort}
\affil[1]{Netherlands Institute for Radio Astronomy (ASTRON)\\Dwingeloo\\the Netherlands}
\affil[2]{Leiden Institute for Advanced Computer Science (LIACS)\\Leiden University\\Leiden\\the Netherlands}
\date{}

\begin{document}



\maketitle


\section{abstract}
Currently, data-intensive scientific applications require vast amounts of compute resources to deliver world-leading science. The climate emergency has made it clear that unlimited use of resources (e.g., energy) for scientific discovery is no longer acceptable. Future computing hardware promises to be much more energy efficient, but without better optimized software this cannot reach its full potential. In this vision paper, we propose a generic AI-driven co-design methodology, using specialized Large Language Models (like ChatGPT), to effectively generate efficient code for emerging computing hardware. We describe how we will validate our methodology with two radio astronomy applications, with sustainability as the key performance indicator.

This paper is a modified version of our accepted SuperCode project proposal. We present it here in this form to introduce the vision behind this project and to disseminate the work in the spirit of Open Science and transparency. An additional aim is to collect feedback, invite potential collaboration partners and use-cases to join the project.

\section{Introduction}
Data-intensive science, such as radio astronomy and high-energy physics, requires vast amounts of compute resources to deliver world-leading science. The current energy and environmental crises drive a strong desire to do science in a manner that minimizes the environmental impact we make while maximizing the science we can deliver. Modern special-purpose compute architectures promise to be much more power efficient than general-purpose systems. However, leveraging these novel architectures is time-consuming and thus expensive due to the effort it takes to port existing code to a new architecture and the increasing complexity and specialization of hardware components. 

In this vision paper, we present \textbf{SuperCode} (SUstainability PER AI-driven CO-DEsign), a novel approach on how to improve effective co-design of hardware and software, since this is essential to ensure that the resulting hardware and software combination is fit for purpose and able to run efficiently. We hypothesize that recent advances in code generation with AI-based Large Language Models (LLMs, e.g., ChatGPT) can be a catalyst for this process. We propose a systematic AI-driven co-design methodology that can drastically reduce the turn-around time to evaluate emerging technologies for data-intensive science, with sustainability as Key Performance Indicator (KPI). To validate our novel approach, we will explore two radio astronomy science cases and investigate their most optimal and sustainable emerging technology platform. With the partners in our project, we will explore opportunities in other domains like climate research, remote sensing and earth observation.

The primary contributions in this paper can be summarized as follows:
\begin{itemize}
\item We present our vision for a novel AI-driven co-design methodology that promises to greatly improve the turn-around time for the evaluation of emerging and new technologies for data-intensive science.
\item  In the spirit of Open Science, team science and accountability, we publish a modified version of the project proposal for the accepted SuperCode project.
\item We define sustainability as first class citizen, and use the methodology we introduced in previous work~\cite{broekema2020optimising} to explicitly reason about this.
\item We provide a comprehensive overview of the current state of the art in code generation with LLMs.
\item Using a simple example we show the current limitations of Large Language Models (LLMs) for co-design applications: current models show promise, but are over-estimating their capabilities.
\end{itemize}

\section{Problem statement}
Data- and compute-intensive science, or eScience, is now firmly established as the fourth science paradigm~\cite{hey_fourth_2009}, as introduced by Tony Hey and Jim Gray. While this opens exciting new abilities to explore new science in vast amounts of data, this comes at very significant processing and energy cost~\cite{centraal_bureau_voor_de_statistiek_elektriciteit_2021,project_lean_2019}. This cost is often overlooked and poorly understood or appreciated. In the current climate crisis we can no longer accept that world-leading science has an unknown, and more importantly, potentially unconstrained environmental impact~\cite{project_lean_2019}. We argue that a careful and deliberate consideration needs to be made whether the scientific impact outweighs the potential environmental impact by the processing required. Optimization and a better mapping of software to hardware can shift this balance in our favour.

\begin{figure}
    \centering
    \includegraphics[]{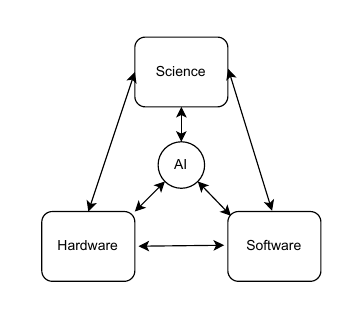}
    \caption{Adding science to the hardware/software co-design loop and leveraging AI to facilitate this process}
    \label{fig:tri-co-design}
\end{figure}

A key part in the optimization process is the ability to efficiently co-design software, hardware, and science.
However, such co-design is time-consuming and complex due to required manual optimisation and the iterative nature of the process. Furthermore, traditional co-design only takes hardware and software into account, missing an opportunity to have emerging technologies drive new scientific discovery. 
By developing and extending novel AI techniques like generative large language models, we will significantly accelerate the co-design loop, informing and challenging the science community to leverage the evolving scientific potential generated by these emerging technologies. 
This project combines two distinct co-design concepts: parallel hardware/software co-design that is accelerated and supported by collaborative human/AI co-design. 
To illustrate this, we offer two examples from radio astronomy, from the LOFAR telescope~\cite{van2013lofar} to be precise. LOFAR is a large, European-scale distributed sensor network containing tens of thousands of receivers that together form one large virtual instrument. It is the largest telescope at this frequency range, producing terabits of data per second. Most of the data processing is done in software.

First, the LOFAR correlator and beamformer, which have gone through several implementations, with differing word-sizes tailored to the hardware characteristics. On BlueGene/L and BlueGene/P we used double precision complex floating point arithmetic, due to the availability of the double hummer FPU, which was built specifically for that purpose~\cite{romein_astronomical_2006,Romein:11a,romein_lofar_2010}. On Cobalt, the GPU-based correlator beamformer currently employed, we use single precision floating point arithmetic, since that is precise enough for the science cases, while doubling the computational performance~\cite{broekema_cobalt_2018,varbanescu2009building,sclocco2012radio,van2010building}. 
More recent GPUs have AI-optimized tensors cores that utilise even smaller word sizes. 
These are experimentally used and may be considered for future implementations of the LOFAR correlator and beamformer~\cite{romein_tensor-core_2021}. 
Smaller word sizes reduce the bandwidth and storage requirements, which can be traded for spectral bandwidth, giving the researchers more scientific flexibility.

Second is the LOFAR polyphase filter, which we made configurable in the number of frequency channels produced~\cite{Romein:11a}. This provided the scientists with a parameter that was not originally in the design requirements. This allowed a trade-off between temporal sensitivity and frequency resolution, enabling better Radio Frequency Interference (RFI) mitigation, while simultaneously supporting higher time resolution for observation modes that need it, such as pulsar search~\cite{van2016towards,van2018real,levin2017pulsar,sclocco2013pulsar,sclocco2014real,sclocco2015finding}. 
Alternatively, smaller data products could be generated, which requires fewer resources to store and process, thus limiting environmental cost. 
These two examples illustrate the potential flexibility gain of software-science co-design. 
The potential energy savings of co-design also are large. 
For example, the LOFAR correlator became approximately 6.6x more power efficient thanks to migrating from a Blue Gene/P supercomputer to a GPU-based cluster (from an estimated 338 to 51 MWh/year), reducing the carbon footprint with approximately 141 tonnes of CO$_2$/year~\cite{kruithof_energy_2023}.

Generative AI, in particular large language models~\cite{wu_autogen_2023,brown_language_2020,wei2022emergent}, have shown a remarkable ability to do mundane coding tasks~\cite{godoy_evaluation_2023}. 
It is now no longer doubtful that computer programming will change dramatically in the coming decade, as demonstrated with tools like GitHub co-pilot~\cite{nguyen_empirical_2022}. 
In this project we introduce AI-drive co-design: we will investigate the ability of AI to accelerate co-design tasks by estimating performance potential of emerging technologies for specific tasks and aiding in the often arduous porting of code. 
It should be noted that the cost of training or specializing such AI is not free of environmental cost, and this should be considered.

\subsection{Illustrative Example}
In the appendix~\ref{sec:appendix} to this paper we show an example of the current state-of-the-art, which demonstrates our vision using existing, not fine-tuned, models. In this case we use QWEN 2.5 Coder 14B Instruct\footnote{\url{https://huggingface.co/Qwen/Qwen2.5-Coder-14B-Instruct-GGUF}}\footnote{Other LLMs were also tested, with essentially the same result}. The result, shown with the context removed and only the generated code reproduced, demonstrates that the concept has merit but that current models are not suitable for the task at hand. The following prompt was used:
\begin{quote}
\texttt{Write me an efficient polyphase filter for the RISC-V architecture with V vector extensions in assembly.}
\end{quote}
The result demonstrates that this model is aware of the RISC-V architecture and uses the correct assembly vector instructions. However, it does not understand what a polyphase filter (essentially a bank of Finite Impulse Response (FIR) filters feeding into a Fast Fourier Transform (FFT)) is. The result seems to use V-extension calls, but the result does not make sense. This illustrates that the current general purpose models clearly needs to be fine-tuned for our applications and that a human-in-the-loop approach is essential for the application we have in mind.

\section{Vision and methodology}

The primary transverse goal of SuperCode is to reduce the environmental impact of data-intensive applications, through a novel approach of AI-driven co-design. 
In contrast to normal co-design, where computational performance or efficiency is used as Key Performance Indicator (KPI), we introduce a sustainability score instead. 
This approach not only benefits our radio astronomy use-cases, but also immediately benefits our commercial partners.

Data-intensive science, and in particular radio astronomy, thrives by virtue of the availability of abundant and affordable computing to process the vast amount of data generated by modern instruments. Current generation instruments produce data at petascale~\cite{broekema_cobalt_2018}, this is expected to increase to exascale in the near future~\cite{van2016towards,broekema2012exascale,broekema2012dome,engbersen2014ska}. Processing such data for primary science cases is already challenging, mining for serendipitous data is likely infeasible due to prohibitive energy costs. The current climate crisis makes it important to visualize and minimize the environmental impact of the processing done. Generative AI, and in particular large language models like ChatGPT, have already shown a remarkable ability to generate code for well-known architectures~\cite{godoy_evaluation_2023,nguyen_empirical_2022,barke_grounded_2023,le_coderl_2022}. We will investigate if we can push the boundaries of these AI models to generate code for more energy efficient emerging hardware platforms. Our method will thus face an extremely challenging problem: AI needs to generate code in a language or paradigm that does not know about. In this project we will develop and test a novel AI-assisted hardware-software-science co-design methodology (see Figure \ref{fig:tri-co-design}).

To validate our approach, we apply this methodology to two relevant and significantly different use-cases in radio astronomy using emerging technologies, with sustainability as the key performance indicator. Terrestrial and space-based radio telescope concepts will be used to exercise the proposed methodology. Both observe similar astronomical structures, but the very different scale, energy budget and environmental conditions result in very different optimal choices. We describe the use-cases in detail in Section \ref{sec:usecases}.

We use radio astronomy as application domain due to its extremely challenging processing requirements. However, we emphasize that the approach is generic and applicable to any code. Using the tools developed and the knowledge gained in this project, we will engage the science community and collaboratively explore novel directions of research made possible by this technology push. The addition of science to the traditional software-hardware co-design duopoly allows emerging technologies to be efficiently leveraged in new and exciting ways and opens potentially unexplored science cases that could not be processed with conventional technologies. We closely collaborate with the Netherlands eScience Center\footnote{\url{https://www.esciencecenter.nl/}} and SURF\footnote{\url{https://www.surf.nl/}} to valorise the approach in other data-intensive research areas (e.g., climate research, remote sensing, earth observation, particle physics). In addition, we have gathered a strong user committee to facilitate industry uptake. This is done primarily by hosting several hackathons, where the participants are invited to apply the research to their own applications and/or architectures. We will proactively share our methodology and results via the user committee, the workshops we will organize, papers, and other dissemination activities.

Three main research questions drive this project, each with several sub-research questions.

\begin{itemize} 
\item[\textbf{RQ1}] \textbf{Can AI-driven co-design make data-intensive applications more sustainable?}
    \begin{itemize} 
    \item[RQ 1.1] Can we structure and describe the state of the art in the use of sustainable AI in software/hardware co-design?
    \item[RQ 1.2] How do we define and quantify sustainability for data-intensive applications? 
    \end{itemize}
\item[\textbf{RQ2}] \textbf{Can trained AI models be generic and flexible enough to model future von Neumann based
architectures as well as non-traditional hardware like neuromorphic chips?}
    \begin{itemize} 
    \item[RQ 2.1] What is the best way to specialize an LLM for co-design? 
    \item[RQ 2.2] Can we use an LLM to translate from CUDA (NVIDIA) to HIP (AMD), and how does this perform? 
    \end{itemize}
\item[\textbf{RQ3}] \textbf{Which emerging technologies are most suitable for data intensive applications?}
    \begin{itemize} 
    \item[RQ 3.1] Which of the investigated emerging technologies is most sustainable for terrestrial radio telescopes?
    \item[RQ 3.2] Which of the investigated emerging technologies is most sustainable for space-based radio telescopes? 
    \end{itemize}
\end{itemize}

We address these research questions in the light of sustainable data-intensive science, meaning that emerging technologies are evaluated for their ability to reduce the environmental impact of scientific discoveries.

\subsection{Sustainability}
Scientific discovery is moving at an unprecedented pace, and many areas of research are limited in their potential by the lack of sufficient signal and data processing capacity~\cite{broekema2012dome}. However, the age of data-driven scientific discovery potentially comes at a very significant environmental impact. Signal processing for the LOFAR telescope, excluding science processing done by the astronomer, exceeds 500~MWh per year~\cite{kruithof_energy_2023}. With increased capabilities this is expected to increase over the next couple of years. Future telescopes, like the Square Kilometre Array (SKA), are scaled such that procuring sufficient compute capacity is initially infeasible~\cite{engbersen2014ska}.  Even the partial system in that design will likely require MW-scale power.  Efforts are underway to gain insights into the environmental impact that groundbreaking research infrastructures have~\cite{kruithof_energy_2023,martin_comprehensive_2022,li_toward_2023}. However, there is currently no measure for the environmental impact of a scientific discovery.

Part of our vision is to visualize the resources required to make science possible.  While energy consumption is the most obvious parameter in this context, it is by no means the only one. This is offset by the science output and/or economic impact, this is the societal value created. The latter was studied for e.g. LOFAR by the Rathenau institute~\cite{tjong_tjin_tai_impact_2019}. While the definition of science value is bound to be controversial, this may be defined in terms of peer-reviewed publications, scientific discoveries, or prestigious prizes. We optimize for relative science value, which we define as the total value created (total value of ownership, TVO) divided by the resources consumed (total cost of ownership, TCO), shown in equation \ref{eq:relative_science_value}.

\begin{equation}
    M_{S} = \frac{TVO}{TCO}
\label{eq:relative_science_value}
\end{equation}

More detail about this method and its application in science can be found in our earlier work~\cite{broekema2020optimising}.  Using this relation, we will design hardware and software combinations that maximize science per unit of environmental impact (the used unit is flexible, and can be energy, CO$_2$, water, etc., or combinations of those). While some resources will inevitably be consumed, we need to be conscious of both the cost of those resources, the value that can be created, and be responsible enough to maximize the science we do with those. Using the proposed methodology, we aim to create a process that provides tailored advice for different science cases.

\subsection{Sustainability as Key Performance Indicator}

In this project we distinguish two separate classes of key performance indicators: at macro and micro level. 
At the micro level, we evaluate the effectiveness of our AI-models by measuring the effort required to evaluate a particular emerging technology for our two use-cases. 
This involves an estimate of the quality of the produced prototype code, i.e., does it work as is, what performance is achieved, and does it produce accurate results, as well as the accuracy of the sustainability estimate produced by the AI model. 
Furthermore, we track the effort required and time needed to evaluate complex emerging technologies. 
We will thus use concrete measurements to validate our hypothesis that AI-driven co-design is able to significantly reduce the effort and time required to adapt and test hardware and software combinations for data-intensive science applications.
At macro level our co-design KPIs focus on sustainability. We aim to minimize the environmental impact of data-intensive science by leveraging emerging technologies to optimize the efficiency of signal processing needed to turn instrument data into scientific data products. Traditionally we would optimize for computational performance or efficiency, in this project we rather focus on sustainability, the exact definition and metric of which is to be defined in the first stage of the project (RQ 1.2, PhD-2). 
Likely this is a combination of scientific potential, energy consumed, time to answer, resources and environmental impact required for production, recycling potential and cost, and others reduced to a sustainability score. 
We note that the environmental impact of technology will be an estimate based on public information and best effort estimates.

We refer to notable efforts by companies like Fairphone to be more transparent with their annual environmental impact and the progress made to reduce this~\cite{noauthor_impact_nodate}. Defining scientific potential of an instrument or technology will be challenging and is bound to be controversial. This will be the focus of the first part of the project but is likely to include potential for peer-reviewed publications, ground-breaking discoveries, instrument mean time between failure and mean time to recovery and ability to host multiple experiments. We acknowledge that this will be difficult to define but it is an important and often forgotten metric. We note that as a fallback scenario value can be defined as a constant, provided we ensure all technologies deliver similar
scientific potential.

\subsection{AI-driven co-design}
Co-design generally uses an existing algorithm and tailors the hardware and software to make sure the implementation is efficient. This process is complex and time consuming, requiring hand-optimizing computational hotspots and a vast amount of domain- and platform-specific knowledge~\cite{varbanescu2009building,sclocco2012radio,van2010building,van2011correlating,sclocco2013pulsar,sclocco2014real,sclocco2016real,sclocco2014auto}. Therefore, it is currently unfeasible to quickly react to changing scientific requirements, nor to discuss unforeseen opportunities with the researchers. Essentially, the scientist currently is not an integral part of the co-design loop. Exploiting the latest AI developments, we aim to turn this around.

Generative AI, and in particular large language models like ChatGPT, have shown a remarkable ability to solve straightforward coding
tasks. Products like GitHub co-pilot\footnote{\url{https://github.com/features/copilot}}, Codeium\footnote{\url{https://codeium.com/}} and BlackBox\footnote{\url{https://www.useblackbox.io/}} are changing the way software is developed~\cite{wu_autogen_2023,godoy_evaluation_2023,nguyen_empirical_2022,barke_grounded_2023,le_coderl_2022}, adding a tireless companion to the programmer that helps to avoid common pitfalls and write the generic code that is often needed. This allows the programmer to focus on functional correctness, performance and scalability~\cite{barke_grounded_2023}. Our hypothesis is that the use of  AI in co-design can drastically reduce the effort needed to evaluate emerging technologies, making their use far more viable. This in turn leads to more sustainable science through more efficient use of energy and/or other resources. While we will test closed models, our approach is generic and we will favour open models like Llama2~\cite{touvron_llama_2023} to ensure reproducibility. We acknowledge that these models do not perform as well as closed models. Llama2 was optimized for dialog applications using reinforcement learning.

\begin{figure}
\centering
\includegraphics[]{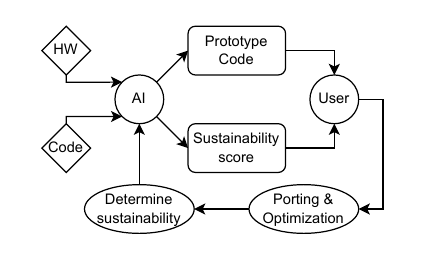}
\caption{A high-level representation of our AI-driven co-design vision}
\label{fig:ai-and-sustainability}
\end{figure}

A well-known property of LLMs is that they can hallucinate~\cite{wei2022emergent,zhang_sirens_2023}, and thus generate unrelated or incorrect output. This is commonly seen as a bug. In this project, we see this as a feature instead, and aim to tap into this creativity to generate novel solutions. I.e., we want to use this emerging knowledge for emerging technologies. By combining AI with the concept of the human in the co-design loop, the creativity exhibited by large language models is both constrained and leveraged.

By evaluating existing generative models and tailoring and testing our own models based on existing foundation models, we will build an AI co-design companion that will assist both programmer and system architect to design a sustainable hardware and software combination. Initially we will prompt an existing AI large language model with a combination of software and a detailed technical description of an emerging technology under investigation. Figure~\ref{fig:ai-and-sustainability} shows a high-level overview of our AI-driven co-design concept. The methodology we envision works as follows.

We will add more information to the foundation models through matrix factorization techniques like low-rank adaptation (LoRA~\cite{hu_lora_2021}, LoHA~\cite{hyeon-woo_fedpara_2021}, LoKr~\cite{edalati_krona_2022}) or other emerging compute-efficient training/fine-tuning techniques (DyLoRA~\cite{valipour_dylora_2023}, GloRA~\cite{chavan_one-for-all_2023} or $(IA)^3$~\cite{liu_few-shot_2022}). This way, we tailor the AI for our needs. We will train these for every emerging technology we investigate, providing the knowledge needed to generate code and estimate the sustainability of the platform, and the two use-cases at the heart of the project. Recent successful work in using LLMs for Chip design indicates that LLMs are indeed capable of grasping new architectures~\cite{liu_chipnemo_2023}. Additionally, we will investigate prompt tailoring and retrieval-augmented generation on modern models~\cite{xu_expertprompting_2023,gao_prompt_2023,lin_ten_2023}, where we take advantage of longer context lengths achieved by newer LLMs that allow ever larger prompts that would ultimately allow us to provide both the entire architecture description as well as the reference code. We will compare these two approaches. These techniques will form the heart of our AI-accelerated co-design process (RQ 2.1, PhD-1). The trained specializations will be publicly released.

We will first train generative foundation models with reference implementations of the scientific data algorithms we use (in our case signal processing algorithms like FIR filter banks~\cite{van2012polyphase}, FFTs, beam forming~\cite{sclocco2012radio}, correlation~\cite{broekema_cobalt_2018,van2010building,van2011correlating,romein_tensor-core_2021}, dedispersion~\cite{sclocco2016real,sclocco2014auto}, etc.). These reference implementations are not optimized for performance, but for explainability and correctness. Using the existing base of open-source radio astronomy code, which is most of the code running current telescopes, we will train our own model. To validate our approach, as a first step we will use LLMs to translate existing Nvidia GPU code to use AMD GPUs. Resulting code and performance will be compared to those generated by the HIPify~\cite{homerding_evaluating_2020} tool provided by AMD, which functions as the ground truth, making this an attractive first step that allows us to validate our methodology in an early stage (RQ 2.2, PhD-2). These ported codes will be publicly released.

Next, we will fine-tune the models with existing hand-optimized implementations, including CPU codes with vector instructions (AVX-512), and GPU codes. We currently use auto-tuning~\cite{sclocco2016real,sclocco2014auto,van_werkhoven_kernel_2019} to generate and test executable code for many different optimization options. In this project, we will feed all these generated codes into the AI model, providing us with sufficient training data. What makes this unique is that we measure the energy efficiency of the generated codes with unprecedented accuracy~\cite{van_werkhoven_kernel_2019}, allowing us to construct a cost function, in turn enabling us to use reinforcement learning to steer the AI towards more energy efficient, and thus potentially more sustainable, solutions~\cite{le_coderl_2022}. Additionally, we can train the model on a host of technical documentation of existing and emerging hardware, for so far as publicly available. The fully trained and fine-tuned trained model will then be tasked to generate sustainability estimates and prototype codes for emerging architectures, based on its learned representations and the new hardware description. The sustainability estimate will be compared to the final measured index. We note here that the field of generative AI is evolving at an unprecedented pace. Any technologies and solutions mentioned in this paper may become obsolete during the SuperCode project. We will monitor the field carefully and use an agile approach to adjust our solutions as needed.

\subsection{Emerging Technologies}
To reduce the environmental impact of data-intensive science, we turn to emerging technologies. The inevitable demise of Moore’s law scaling has given rise to a host of alternative technologies that aim to offer improved performance and efficiency at lower cost. One of the earliest examples and one that is now firmly in the mainstream, is General Purpose computing on Graphics Processing Units (GP-GPU)~\cite{madougou2016landscape,madougou2014empirical}. Many techniques leverage specialization, where special purpose components perform specific tasks more efficiently than general purpose hardware. This is very costly due to non-recurring engineering expenses, and data-intensive science is not expected to procure sufficient volume to recoup that investment, even at the scale of the SKA. However, we do have the opportunity to re-use specialized hardware tailored for other applications, AI currently being the most obvious specialized hardware target. Examples are the tensor core correlator that leverages AI-focused cores on modern GPUs to improve efficiency over older GPUs by several factors~\cite{romein_tensor-core_2021}\footnote{\url{https://git.astron.nl/RD/tensor-core-correlator}}. 

In the short term we will target specialized cores on proven hardware solutions for use in our use-cases. Specifically, AVX-512 extensions in CPUs~\cite{carneiro_lightweight_2021}, tensor cores in NVIDIA GPUs, and matrix core engines on AMD GPUs. This will validate our approach, while simultaneously reducing the risk for the PhD candidates. Next, we will apply the approach on emerging but conventional silicon-based hardware platforms, like the planned accelerators developed in the European Processor Initiative (EPI), the Intel Ponte Veccio accelerator, and currently not publicly described products by startup companies like NextSilicon\footnote{\url{https://www.nextsilicon.com/}}. SURF’s open innovation lab\footnote{\url{https://www.surf.nl/lab}}, and in some cases DAS-6~\cite{bal_distributed_2000} or its successor will enable access to these emerging architectures. Finally, we will test our approach on more esoteric emerging technologies, such as neuromorphic or memristor based systems, without material changes to the underlying methodology. This will test our hypothesis that the AI-driven approach should be flexible enough to accommodate such very different platforms transparently, greatly reducing the effort and turnaround needed to evaluate such systems for our use-cases.

\section{Use-cases}
\label{sec:usecases}
We will apply our methodology to two use-cases: one terrestrial and one space-based. We have selected these because they are both challenging but put vastly different constraints and requirements on the processing platform. Moreover, compared to data-intensive applications in general, these are at extreme ends of the scale. If we have validated these, it is plausible that our methodology also is applicable for less data intensive applications.

\subsection{Use-case 1: Terrestrial large-scale distributed radio telescopes}

\begin{figure}
    \centering
    \includegraphics[width=\columnwidth]{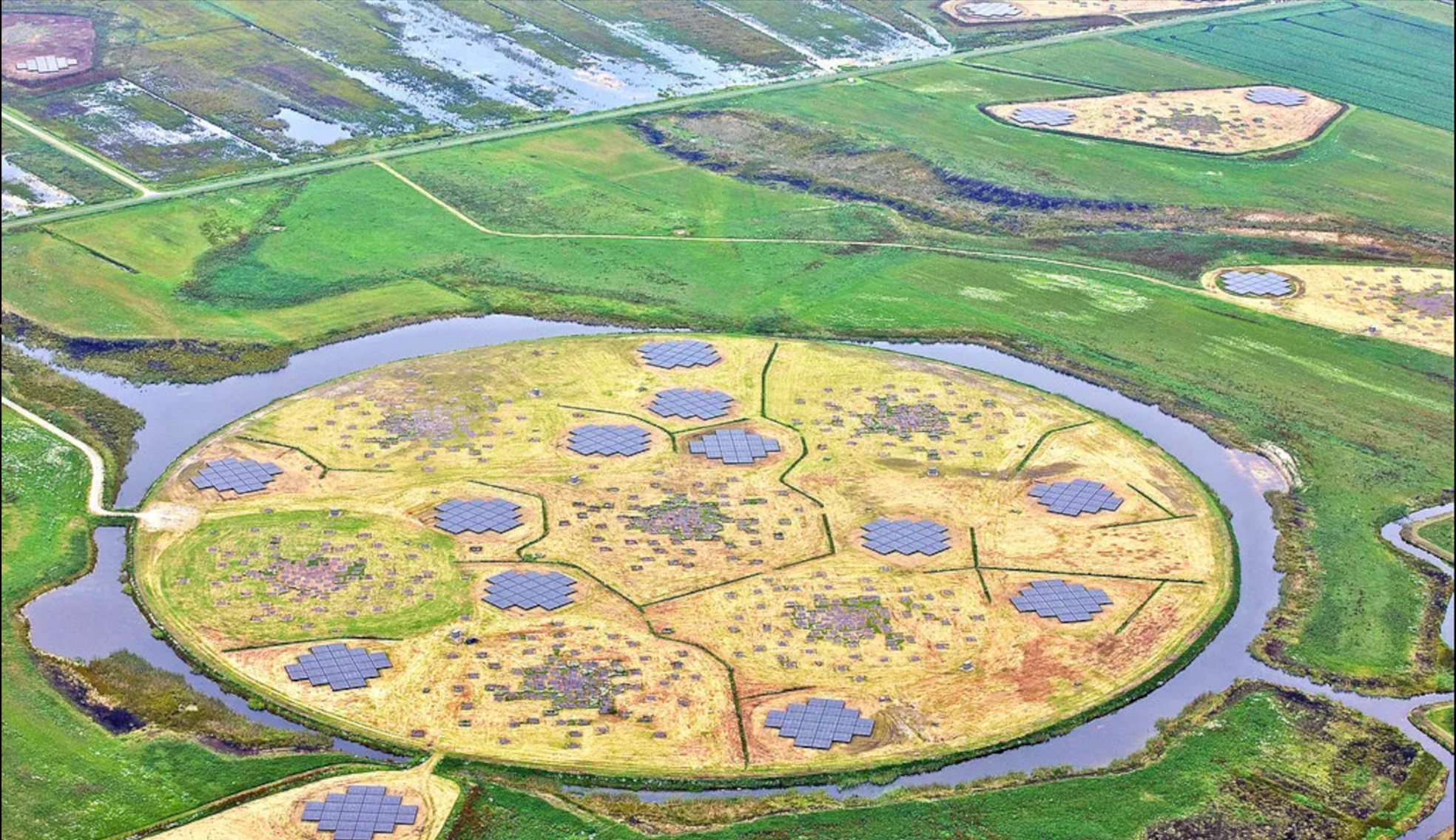}
    \caption{The core of the LOFAR telescope.}
    \label{fig:LOFAR}
\end{figure}

Our first use-case focuses on terrestrial large-scale distributed radio aperture synthesis arrays. These are at the forefront of low- and mid-frequency radio astronomy. Specifically, we will use The LOw Frequency ARray (LOFAR)~\cite{van2013lofar}, designed and built by ASTRON (See also Figure~\ref{fig:LOFAR}), and Square Kilometre Array (SKA)~\cite{schilizzi2008square,broekema2015square} telescopes, currently under construction by a multi-national consortium including ASTRON. Most partners in our user committee are also involved in the construction of the software pipelines of the SKA.

Aperture synthesis arrays create a virtual telescope by combining multiple geographically separated sensors. These all sample the electromagnetic spectrum that, according to the Van Cittert-Zernike theorem~\cite{zernike_concept_1938} can be considered to come from the same distant source. Correlating many combinations of sensors gives us a sparse set of points that have a Fourier-relation with the sky image. We generally take advantage of the earth’s rotation as well as the coherent nature of the signal of interest over incoherent noise to fill in the sparse image and amplify the weak astronomical signals. Radio astronomy requires a lot of signal processing to turn what is essentially noise into a scientific data product. The computational requirements for such instruments scale dramatically ($\mathcal{O}(n^3$) - $\mathcal{O}(n^4)$) with the size of the telescope, both due to increased data volumes and more processing required per unit of data, which is also a key driving factor in sensitivity and resolution. 

LOFAR, a state-of-the-art low-frequency array, requires processing at tera-scale. The SKA, which is about an order of magnitude larger in terms of receivers, is expected to require peta-scale processing~\cite{broekema2015square}. The initial procurement of compute infrastructure will not cover that requirement, mostly due to budget and energy constraints. The scientific potential of current and future radio telescopes is limited by the availability of sufficient affordable data processing capacity and software. In use-case 1, we will evaluate the sustainability aspects of signal processing algorithms for terrestrial telescopes, like FIR filter banks~\cite{van2012polyphase}, FFTs, beam forming~\cite{sclocco2012radio}, correlation~\cite{broekema_cobalt_2018,van2010building,van2011correlating,romein_tensor-core_2021}, dedispersion~\cite{sclocco2016real,sclocco2014auto}, etc.

\subsection{Use-case 2: Space-based radio telescopes}

\begin{figure}
    \centering
    \includegraphics[width=\columnwidth]{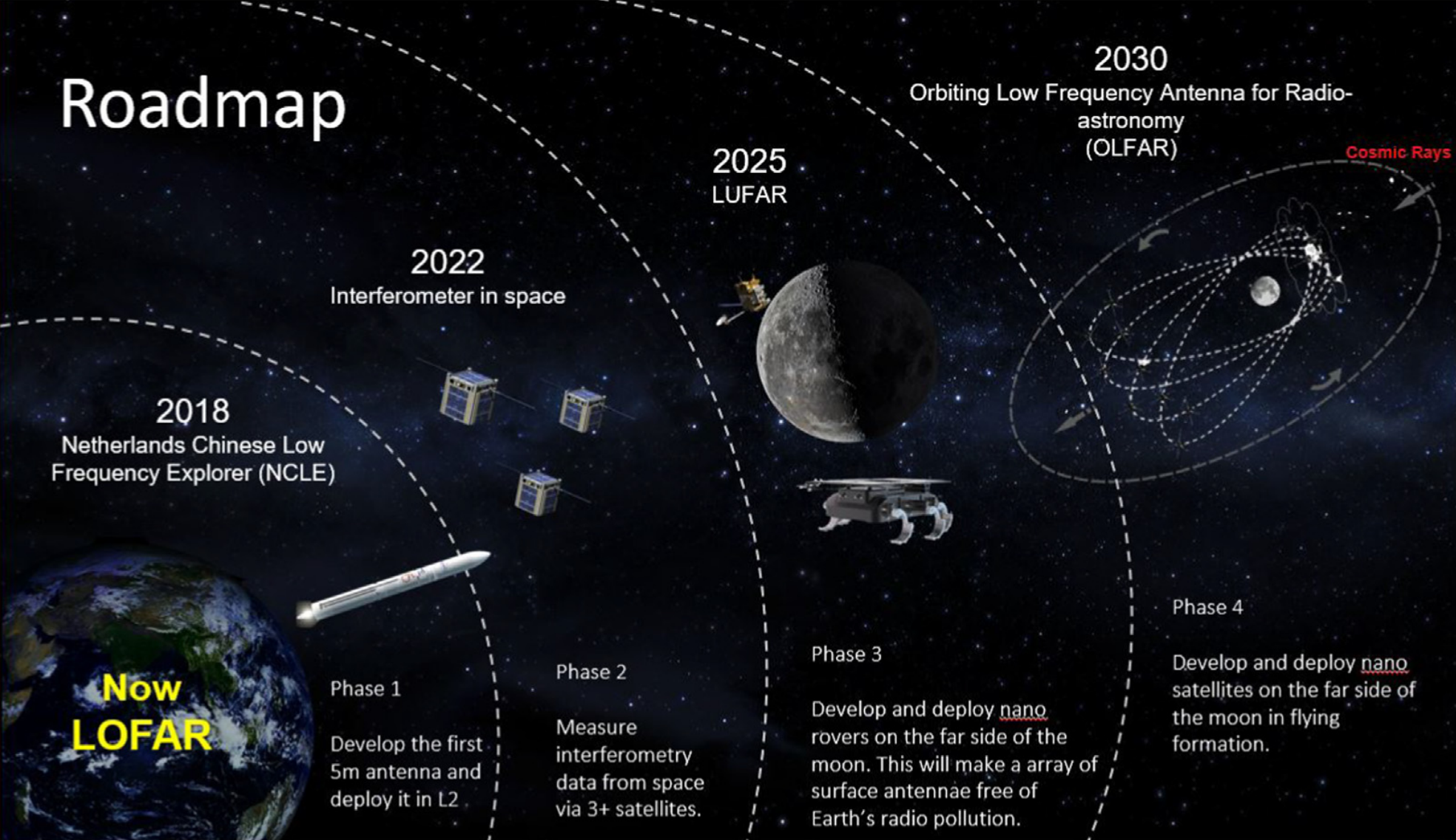}
    \caption{The OLFAR roadmap, picture courtesy \cite{bentum_roadmap_2020}.}
    \label{fig:OLFAR}
\end{figure}

Use-case 2 is more speculative. The promise of higher sensitivity thanks to less ionospheric disturbance and Radio Frequency Interference (RFI)~\cite{van2018real} and an unexplored low-frequency band due to the opaqueness of the atmosphere to frequencies around 10MHz, is generating interest for space-based interferometers. Such instruments include both orbiting swarms (OLFAR~\cite{van_vugt_calibration_2016,bentum_roadmap_2020}, ALO~\cite{jester_science_2009} and DSL~\cite{chen_discovering_2020}, as well as lunar surface instruments (Farside~\cite{burns_lunar_2021} and LuSEE’Night’~\cite{bale_lusee_2023}). The OLFAR roadmap is presented in Figure~\ref{fig:OLFAR}. The hostile extra-terrestrial environment and limited available energy, volume, mass and ability to dissipate heat, lead to unique and quite different concepts and challenges~\cite{rajan_space-based_2016} compared to terrestrial telescopes, making this an interesting use-case.

Where terrestrial radio telescopes favour the transport of as much data as feasible to a central location for processing, this is not the case for space-based telescopes. Here, due to excessive bandwidth constraints and costs, edge processing is key. Furthermore, in space, energy is not abundant and heat dissipation of signal processing systems may be challenging. Cosmic radiation requires space-hardened systems, which are based on older (i.e., with larger gate sizes) production technologies, whereas from a sustainability viewpoint we would favour the newest processes. Note that these considerations are immediately applicable to space-based earth observation, where similar constraints are encountered. Project partners Sioux and S[\&]T have extensive experience in earth observation and have expressed a keen interest in this similarity.

The signal processing is similar for space-based radio interferometry. That said, sustainability will be very different, considering energy in space is sustainably generated. Thus, production and launch will be much more dominant in the equation. Furthermore, we need to consider the environmental impact after the useful life of the instrument. If in orbit, this means de-orbiting to avoid creating space junk. In use-case 2, we will evaluate the which emerging architectures are most suitable and sustainable for space-based telescopes.

\section{Scientific and Societal Impact}
We firmly believe that the use of LLMs will fundamentally and disruptively change the way both academia and industry develop software. The role of software developers will not disappear but will require developers to use and understand this novel AI technology to be more productive and efficient. The Supercode project has six industrial partners who  already collaborate with us in designing and constructing software pipelines for the SKA. The partners were carefully selected on what is their core business: software development in business and industrial applications. Therefore, they will be very much affected by LLM-based code generation. By using the technologies developed in this project in their own use-cases during the workshops we will organize, they will be better prepared for the future, thus significantly enlarging the societal impact of our research.

To ensure that our methods and results are transferred to our industrial partners we will organize workshops and hackathons, with their broader organizations and stakeholders beyond the user committee. These workshops are broader in terms of the audience and also include the software development teams constructing the SKA software pipelines. In these workshops we will discuss our results, show demos, but will also take a hands-on approach to tackle concrete use cases brought in by the participants. The advantage of our methodology is that it enables quick prototyping and experimentation, allowing the users to quickly get a feel for what is possible. Towards the end of the project, we will develop open training materials in Software Carpentry style, enabling an even broader community to implement the methods we will develop. This ensures long lasting impact, also after the project has finished.

We will develop tools and instrumentation to maximize science output per unit of environmental impact. The sustainability score, the KPI in this project, is a trade-off of value (science output) and cost (environmental impact), clearly indicating that societal and scientific impact are of comparable focus. The energy usage of data centres, often to a large extent caused by AI applications, is unsustainable, and will have to be reduced. By not training full models, but instead using and finetuning and extending existing pruned foundation models, we aim to show that much more efficient approaches are feasible, in science, but also in industrial applications.

By implementing our work in concrete use cases related to the SKA, and the direct involvement of the SKA software developer teams, we embed our work in the SKA community. We will publish the results of this project in top conferences and journals. Because of the interdisciplinary nature of the work, we aim to publish in both computer science and astronomy. We take the approach of first publishing extended abstracts in astronomy, mostly to disseminate and validate the work with our peers. Next, we publish the methodological aspects in top computer science venues (e.g., supercomputing, PASC, IPDPS, ICS). Finally, we publish the applied results in astronomy journals (A\&A, MNRAS, astronomy \& computing, etc.). With this triple-target publication strategy, we reach different audiences, maximizing our scientific impact. 

\section{Related work}
AI-supported code generation has recently become possible~\cite{austin_program_2021,wang_codet5_2021} and has seen only very limited application in high-performance computing. So far, no research has been done on generating and porting code to emerging architectures. This is challenging since there are no concrete code examples on the emerging architecture, so the LLMs must effectively do transfer learning to port existing codes to new architectures. LLMs have not been used with the specific focus of generating code for exploring more energy efficient combinations of hardware architecture and software implementation.

Godoy et al.~\cite{godoy_evaluation_2023} evaluate AI-assisted generative capabilities on several numerical HPC kernels. They generated codes for a variety of programming models and languages, including C++/OpenMP, OpenACC, Kokkos, SyCL, CUDA, Fortran and Julia. However, all algorithms tested were well known (e.g., the models were trained on many examples), and they did not generated code for emerging architectures. Nevertheless, this does indicate that our proposed approach is feasible.

Currently, LLMs do not understand program semantics, and offer no guarantees about quality of the generated code. Jain et al.~\cite{jain_jigsaw_2022} demonstrated that augmenting LLMs with syntax and semantics-aware program analysis and user feedback improves the output. The interactive properties of LLM models (via the prompt) have not been used for co-design research in this context. Our project will, for the first time, apply the emerging abilities of generative AI to reduce the challenges that are faced in effective co-design. 

IBM is currently using generative AI to modernize legacy applications~\cite{noauthor_harnessing_nodate}. Their work focuses mostly on translation from COBOL to Java, while our approach translates to different architectures. We must translate between programming languages, but also need to exploit different forms of parallelism (semi-)automatically.

EcoOptiGen~\cite{wang_cost-effective_2023} is a hyperparameter optimization framework for LLMs. EcoOptiGen leverages cost-based pruning to reduce the (energy) cost of LLMs. Like EcoOptiGen, we will also explore the AutoGen~\cite{wu_autogen_2023} framework. While EcoOptiGen focuses on inference, our work focuses on training and finetuning aspects. 

CodeRL~\cite{le_coderl_2022} is a framework for program synthesis through pre-trained LLMs and deep reinforcement learning. It uses a critical sampling strategy to automatically generate programs based on feedback from example unit tests. Our envisioned approach extends this with full reference codes and codes for other architectures. Moreover, our philosophy is (semi)-supervised and based around co-design, keeping an expert in the loop.

\section{Conclusions}
In this paper we have presented a novel AI-driven co-design concept. We argue that Large Language Models can significantly accelerate the evaluation and adoption of efficient new and emerging technologies for data-intensive science. By adopting a human-in-the-loop approach we aim to leverage the creativity exhibited by modern LLMs and turn a conceived weakness (commonly referred to as hallucination) into a strength. Our concept will be applied to some of the most challenging data-intensive science use-cases in radio astronomy, validating and exercising the approach to its limit.

The vision presented in this paper will be further explored in the SuperCode project, starting in January~2025.

\section*{Acknowledgements}
This work was funded in 2024, by the Open Technology Programme call of the Technical and Applied Sciences department of the Netherlands Research Council NWO, under grant number 21356. We would like to thank the following partner organizations in the SuperCode project for their support, in-kind funding and valuable discussions: SURF (Raymond Oonk), CGI (Alexandra Zevenbergen), TriOpSys (Joy Ong), [S\&T] (Erik van Mulligen), Netherlands eScience Center (Patrick Bos), and Sioux Technologies (Bas van der Linden). Finally, we would like to thank the SuperCode advisory committee (Mark Bentum, Aske Plaat, Suzan Verberne, Ben van Werkhoven and Albert-Jan Boonstra) for their insightful feedback. 

\bibliographystyle{plainnat}
\bibliography{SuperCode}

\newpage
\onecolumn
\section{Appendix A: Generated (incorrect) Code for a Polyphase Filter.}
\label{sec:appendix}

This appendix shows an example of generated (incorrect!) code for a polyphase filter in vectorized RISC-V assembly using the QWEN 2.5 Coder 14B Instruct model. Note that only the generated code is shown, not the generated explanation and context.
 \\
 
\begin{scriptsize}
\begin{lstlisting}
```assembly
.section .data
input:  .word 1, 2, 3, 4, 5, 6, 7, 8  # Example input data
coeff1: .word 0.5, 0.5, 0.5, 0.5    # Coefficients for subfilter 1
coeff2: .word 0.5, 0.5, 0.5, 0.5    # Coefficients for subfilter 2
output: .space 8                      # Space for output data

.section .text
.globl _start
_start:
    # Load base addresses and lengths
    la t0, input                      # Load input base address
    la t1, coeff1                     # Load coefficient 1 base address
    la t2, coeff2                     # Load coefficient 2 base address
    la t3, output                     # Load output base address
    li t4, 4                          # Length of input (number of elements)

    # Initialize pointers
    add t5, t0, t4, slliw # t5 = input + length * 4 (end of input)
    add t6, t3, t4, slliw # t6 = output + length * 4 (end of output)

    # Loop over input data
loop:
    # Check if we've reached the end of input
    beq t0, t5, end

    # Load 4 elements from input
    vlw.v v0, (t0)

    # Load 4 coefficients from coeff1 and coeff2
    vlw.v v1, (t1)
    vlw.v v2, (t2)

    # Multiply and add for subfilter 1
    vfma.vv v3, v0, v1

    # Multiply and add for subfilter 2
    vfma.vv v4, v0, v2

    # Store results to output
    vsw.v v3, (t3)
    vsw.v v4, (t3)

    # Move to next set of coefficients
    addi t1, t1, 16
    addi t2, t2, 16

    # Move to next set of input and output
    addi t0, t0, 16
    addi t3, t3, 8

    # Repeat loop
    j loop

end:
    # Exit program (system call for Linux)
    li a7, 93
    ecall
```
\end{lstlisting}
\end{scriptsize}

\end{document}